\documentclass[prb,aps,twocolumn,showpacs,10pt]{revtex4-1}
\usepackage{amsmath,amssymb,bm,graphicx,dsfont,color}
\usepackage[latin9]{inputenc}
\usepackage{amsmath}
\usepackage{slashed}
\usepackage{dsfont}
\usepackage{amstext}
\usepackage{amssymb}
\usepackage{amsbsy}
\usepackage{amsthm}
\usepackage{epsfig}
\usepackage{graphicx}
\usepackage{bbm}
\usepackage{hyperref}
\usepackage{color}
 
\begin{document}


\title{Boson topological insulators: A window into highly entangled quantum phases}
\author{Chong Wang and T. Senthil}
\affiliation{Department of Physics, Massachusetts Institute of Technology,
Cambridge, MA 02139, USA}
 \date{\today}
\begin{abstract}
We study several aspects of the realization of global symmetries in highly entangled phases of quantum matter. Examples include gapped topological ordered phases, gapless quantum spin liquids and non-fermi liquid phases. An insightful window into such phases is provided by recent developments in the theory of {\em short ranged entangled} Symmetry Protected Topological (SPT) phases .  First they generate useful no-go constraints on how global symmetry may be implemented in a highly entangled phase.  Possible symmetry implementation in gapped topological phases and some proposed gapless spin/bose liquids are examined in this light.  We show that some previously proposed spin liquid states for $2d$ quantum magnets do not in fact have consistent symmetry implementation unless they occur as the surface of a $3d$ SPT phase.  A second SPT-based insight into highly entangled states is the development of a view point of such states as SPT phases of one of the emergent excitations. We describe this in the specific context of time reversal symmetric $3d$ $U(1)$ quantum spin liquids with an emergent photon. Different such spin liquids are shown to be equivalent to different SPT insulating phases of the emergent monopole excitation of such phases.  The highly entangled states also in turn enrich  our understanding of SPT phases. We use the insights obtained from our results to provide an explicit construction of bosonic SPT phases in $3d$ in a system of coupled layers. This includes construction of a time reversal symmetric SPT state that is not currently part of the cohomology classification of such states.

\end{abstract}
\newcommand{\be}{\begin{equation}}
\newcommand{\ee}{\end{equation}}
\newcommand{\bea}{\begin{eqnarray}}
\newcommand{\eea}{\end{eqnarray}}
\newcommand{\p}{\partial}
\newcommand{\lp}{\left(}
\newcommand{\rp}{\right)}
\maketitle

A focus of modern quantum condensed matter physics is the study of phases of matter whose characterization is not captured by the concepts of broken symmetry and associated Landau order parameters. Striking examples are  gapped phases of matter with topological order. These are characterized by emergent excitations with unusual quantum statistics and ground state degeneracies that depend on the topology of the underlying manifold\cite{Wenbook}. Other examples are gapless phases of matter where the gaplessness is protected but not by a broken symmetry. The most familiar example of such a phase is a Fermi liquid but gapless spin liquids  and various non-fermi liquid phases provide other examples.  A common characterization of these different phases is the presence of non-local many body quantum entanglement in their ground state wave function.  Such phases have come to be known as ``highly entangled phases" of matter. 

A simpler example of phases of matter that are not captured by notions of broken symmetry are the celebrated electronic topological insulators\cite{TIs}. These phases are gapped in the bulk and are {\em short ranged entangled} but nevertheless are distinguished from trivial electronic band insulators. Symmetry plays a key role in maintaining the distinction between these 
different short ranged entangled phases. The electronic topological insulators are well described by free
fermion models that have non-trivial surface states protected by global symmetries. They clearly do not have bulk topological order or fractionalization. The free fermion topological insulator is a member of a more general class of phases - dubbed Symmetry Protected Topological (SPT) phases - which all have no bulk topological order but have protected surface states.  A well known example is the Haldane spin chain in one dimension. A general formal classification of such phases in diverse dimension has been proposed\cite{chencoho2011,kunpub}. Very recently their physical properties in both two\cite{luav2012,tsml,liuxgw} and three dimensions\cite{avts12,xuts13} have been elaborated

The story of the electronic/bosonic topological insulators raises the important question of the role that symmetry plays in the characterization of highly entangled phases.     In the specific context of gapped highly entangled phases the interplay of symmetry and topological order has  recently attracted renewed interest.   For gapless phases/critical points global symmetry obviously plays a much more important role that is very poorly understood. 
In this paper we will study several aspects of the realization of symmetry in these exotic phases - both gapped and gapless, and in both two and three space dimensions. 
Of particular importance to us are the results of Ref. \onlinecite{avts12} on the
protected surface states of three dimensional bosonic SPT phases. The surface phase diagram was argued to admit a phase with surface topological order though the bulk itself has no such order. Furthermore this surface topological order implements the defining global symmetry in a manner not allowed in strictly two dimensional systems. 

We are thus lead to consider in detail consistent implementation of global symmetries in several highly entangled quantum phases. 
First we obtain several new results and insights into both gapped and gapless phases that are allowed to exist in strictly 2d systems. These results have immediate application to theories of quantum spin liquid insulators and of non-fermi liquid metals. Along the way we also obtain an explicit construction of the various 3d symmetry protected topological insulators of bosons studied recently in Ref. \onlinecite{avts12}. In particular we construct a time reversal symmetric $3d$ SPT phase that was suggested to exist in Ref. \onlinecite{avts12} but is not currently part of the cohomology classification of Ref. \onlinecite{chencoho2011}. 

Second we study symmetry realization in three dimensional gapless quantum spin liquids with an emergent photon.  Focusing on time reversal symmetry and on phases that can exist in strictly $3d$ systems we show that different such spin liquids may be distinguished by whether the emergent electric charge excitation is a Kramers singlet/doublet and its statistics. 
We show that this distinction is nicely captured by viewing these phases as different SPT insulators of the dual `magnetic' particle (the monopole).

Ref. \onlinecite{hermele} proposed a formal classification of two dimensional topological order described by  a deconfined $Z_2$ gauge theory in the presence of global symmetries. The topological quasiparticles can in principle carry fractional quantum numbers of the global symmetry. More formally this means that they are allowed to transform projectively under the global symmetry group. The approach of Ref. \onlinecite{hermele} involves finding all consistent ways of assigning projective representations to the different topological quasiparticles. A different classification has also appeared\cite{ran12} that considers  topological order with unitary symmetry but restricts to phases where one of the bosonic quasiparticles has trivial global quantum numbers. 
Earlier Refs. \onlinecite{Santos,Levin} classified all two dimensional time reversal invariant gapped abelian insulators using a Chern-Simons/K-matrix approach. It is expected that all such insulators can always be described by a multicomponent Chern-Simons theory. The key idea of Refs. \onlinecite{Santos,Levin} is that the bulk two dimensional theory can be completely characterized by studying the $1+1$ dimensional edge theory at the interface with the vacuum (or equivalently a topologically trivial gapped insulator). This is a multi-component Luttinger liquid theory  in which operators corresponding to various bulk quasiparticles can be easily identified. In particular constraints coming from global symmetries can be straightforwardly implemented. This approach has recently been used to study other symmetry enriched $2d$ topological order in Refs. \onlinecite{luav13,janet13}. 

 How does the interplay of symmetry and topological order that is only allowed at the surface of 3d systems fit in with the emerging results on the classification of 2d topological order with symmetry? In the first part of this paper we address this question in detail for a few examples. Specifically we restrict attention to boson systems with a few simple internal global symmetries. We also restrict to topological order described by a deconfined $Z_2$ gauge theory. We first 
use the procedure of Ref. \onlinecite{hermele} to obtain all distinct allowed implementation of the global internal symmetry.  Some of these can be realized at the surface of 3d SPT phases. Then we use elementary arguments and the results of Ref. \onlinecite{Levin} to determine which ones of the phases are allowed in strictly 2d systems.  Interestingly the remaining phases are {\em  all} shown to be realized at the surface of 3d SPT phases. 

Why does the Chern-Simons/edge theory approach select out only those phases that can exist in strict 2d while the approach of Ref. \onlinecite{hermele} does not? 
The key point is that the former approach assumes that the state in question can have a physical edge with the vacuum (equivalently a topologically trivial gapped insulator) while preserving the symmetry. For topological order realized at the surface of a 3d SPT phase this possibility simply does not exist. A trivial gapped symmetry preserving state to which the surface topological ordered state can have an interface is forbidden at the SPT surface. In contrast the methods of Ref. \onlinecite{hermele} only worry about consistent assignment of symmetries to the various topological quasiparticles. The requirement that the state allow a physical edge to the vacuum is not part of the considerations of this method. 

Below we will flesh all this out in several concrete examples.  We first study $2d$ gapped $Z_2$ topological order with a few different symmetries in Sections. \ref{bu1rtz2}, \ref{z2t} and \ref{beyondcoh}. Next we use the insights from these results to provide an explicit construction of SPT  phases with the same symmetries in a system of coupled layers in Sec. \ref{clayer}.  We provide a brief discussion of the relationship between the possible surface topological order in a 3d SPT and its bulk topological field theory in Section \ref{rbtft}. 
We turn our attention then to highly entangled gapless phases. In Section. \ref{avl} we argue  that a previously proposed gapless vortex fluid (dubbed the `Algebraic Vortex Liquid') cannot exist 
with time reversal symmetry in strictly $2d$ systems. Section \ref{3du1} contains our results on $3d$ $U(1)$ quantum liquids. We conclude in Section \ref{disc} with a discussion. 
Two Appendices contain important details. In particular in Appendix \ref{duallgw} we describe the surface Landau-Ginzburg theories for the 3d SPT phases of interest in terms of dual vortices with non-trivial structure and discuss the surface phase structure. We also provide an explicit derivation of this dual surface  vortex theory. 

\section{Topological ordered boson insulators: Symmetry $U(1) \rtimes Z_2^T$}
\label{bu1rtz2}

We begin by considering a system of bosons with a global $U(1)$ symmetry and time reversal ($Z_2^T$). The bosons are taken to have charge $1$ under the global $U(1)$ symmetry. In this section we assume that  the boson destruction 
operator $b \rightarrow b$ under $Z_2^T$. This means that the global symmetry group is $U(1) \rtimes Z_2^T$.  We will assume that the topological order in question has 2 non-trivial bosonic particles (dubbed e and m) and a fermion (dubbed $\epsilon$). Any two of these are mutual semions. Further any one of these may be thought of as a bound state of the other two. This corresponds precisely to the excitation structure of a deconfined $Z_2$ gauge theory in two space dimensions. What are the allowed topological phases with $Z_2$ gauge structure according to the analysis of Ref. \onlinecite{hermele}? The time reversal operation ${\cal T}$ when it acts on physical states of the bosons must satisfy ${\cal T}^2 = 1$. Let us denote by ${\cal T}_{e,m}$ the action of time reversal on the e and m particles. The only restriction on these is that they satisfy
\begin{eqnarray}
{\cal T}_e^2 & = & \mu_e \\
{\cal T}_m^2 & = & \mu_m
\end{eqnarray}
with $\mu_{e,m} = \pm 1$.  A value $-1$ of either of these means that the corresponding particle forms a Kramers doublet. 
What about symmetry under global $U(1)$ rotations? Here the distinct possibilities correspond to whether the $(e,m)$ particles carry integer or fractional charge. In the latter case their charge must be shifted from an integer by $\frac{1}{2}$. These possibilities are nicely distinguished by asking about the action of a $2\pi$ global $U(1)$ rotation $R_{2\pi}$. On physical states $R_{2\pi} = 1$. Let us again denote by $R_{2\pi}^{e,m}$ the action on the e and m sectors. We then have 
 \begin{eqnarray}
R_{2\pi}^e & = & \sigma_e \\
R_{2\pi}^m & = & \sigma_m
\end{eqnarray}
with $\sigma_{e,m} = \pm 1$.  The realization of the symmetry in this topologically ordered state is thus described by the numbers $(\sigma_e, \mu_e, \sigma_m, \mu_m)$. Naively this gives 16 phases but we must remember that interchanging e and m does not produce a new phase. This removes 6 possibilities so we are left with a total of 10 phases for this symmetry.

In Table \ref{u1rtz2t} we display the quantum numbers of the e and m excitations of these 10 phases. We label these phases by the excitations that carry non-trivial charge (C) or time reversal (T) quantum numbers. Thus $e0m0$ means both the e and m particles carry trivial quantum numbers, while $mT$ refers to a phase where the m particle is  Kramers doublet and neither e nor m cary half-integer charge, etc. 

 \begin{table}[htdp]
\begin{center}
\begin{tabular}{|c|c|c|c|c|c|}
\hline
Phase & $\sigma_e$ & $\sigma_m$ & $\mu_e$ &$ \mu_m$ & Comments \\ \hline
$e0m0$ & $1$ & $1$ &  $1$ & $1$ & No fractionalization \\ \hline
$eT$ & $ 1$ & $1$ & $-1$ & $1$ & No fractional charge but Kramers \\ \hline
$eC$ &  $-1$ & $1$ &  $1$ & $1$ & $b = \Phi^2$ \\ \hline
$eCT$ &  $-1$ & $1$ & $-1$ &  $1$ & $b = \epsilon_{\alpha\beta} f_{\alpha} f_{\beta}$;  $f_\alpha$ in trivial band  \\  \hline
$eCmT$ & $ -1$ & $1$ & $1$ & $-1$ & $b = \epsilon_{\alpha\beta} f_{\alpha} f_{\beta}$;$f_\alpha$ in  topological band  \\ \hline
\hline
$eTmT$ & $1$ & $1$ &  $-1$ & $-1$ & 3d SPT surface \\ \hline
$eCmC$ & $ -1$ & $-1$ & $1$ & $1$ & 3d SPT surface \\ \hline
$eCTmC$ &  $-1$ & $-1$ &  $-1$ & $1$ &   $ eCmC \oplus eCmT$ \\ \hline
$eCTmT$ & $ -1$ & $1$ & $-1$ & $-1$ &   $ eTmT \oplus eCmT$  \\  \hline
$eCTmCT$ &  $-1$ & $-1$ & $-1$ &  $-1$ &   $ eCTmC \oplus eCTmT$   \\
\hline
\end{tabular}
\end{center}
\caption{ Symmetry action ($U(1) \rtimes Z_2^T$) for  $Z_2$ topological ordered states. The first 5 are allowed in strict 2d while the last 5 can only be realized at surface of 3d SPT phases (or derived from them)}
\label{u1rtz2t}
\end{table}%

Let us now switch gears and consider the possibilities for $Z_2$ topological order with the same symmetry as above but within the Chern-Simons/edge theory approach  of Ref. \onlinecite{Levin}. This will enable us to decide which of the 10 states in Table \ref{u1rtz2t}  can be realized in strictly 2d systems. We will show that only the first 5 of these are captured in the Chern-Simons approach.

In the Chern-Simons description of an abelian two dimensional insulator  the effective Lagrangian is given by a multi-component Chern-Simons term
\be
L=\frac{K_{IJ}}{4\pi}\epsilon^{\mu\nu\lambda}a_\mu^I\p_\nu a_\lambda^J+\frac{1}{2\pi}\tau_I\epsilon^{\mu\nu\lambda}A_\mu\p_\nu a^I_\lambda
\ee
where the current density of quasiparticle $I$ is given by $j^I_\mu=\frac{\epsilon^{\mu\nu\lambda}\p_\nu a_\lambda^I}{2\pi}$. The $K$-matrix gives the topological information of the system, while the charge vector $\tau_I$ is an integer valued charge of each quasi-particle through coupling with the external gauge field $A_\mu$.     The allowed quasiparticles carry integer charge under the different gauge fields $a^I$ which can be expressed in terms of an integer valued vector $l$. 
The mutual statistics of two quasiparticles labeled by $l$ and $l'$ is $\theta_{ll'} = 2\pi l^T K^{-1} l$ while the self-statistics of a quasiparticle is $\theta_l = \pi l^T K^{-1} l$. To describe  $Z_2$ topological order we begin with a $2 \times 2$ K-matrix 
\be
K=\lp\begin{array}{cc}
          0 & 2 \\
          2 & 0 \end{array}\rp
          \label{z2k}
\ee
which captures the statistics of the $e$ and $m$ particles. We will determine the distinct ways in which the $U(1) \rtimes Z_2^T$ symmetry can be realized. First of all note that the electrical Hall conductivity is given by $\sigma_{xy} = \tau^T K^{-1} \tau = \tau_1 \tau_2$. Thus time reversal invariant states must necessarily have at most one of $\tau_{1,2} \neq 0$. 
Henceforth without loss of generality we will therefore set $\tau_1 = 0$ and $\tau_2 = t$.  Next the physical charge of a quasiparticle labeled by $l$ is given by $q_l = l^t K^{-1} \tau
= \frac{l_1t}{2}$. Since we only want to distinguish half-integer physical charge from integer the distinct possibilities correspond to $t = 0,1$. 
Let us now demand time reversal invariance of the Chern-Simons Lagrangian. The symmetry realizations classified by the first approach above assume that the symmetry transformation does not interchange $e$ and $m$ particles. Therefore we restrict attention to that subclass here. For the first term to be time reversal invariant it must be that the spatial components $a_{1i}, a_{2i}$ transform oppositely under time reversal. Further if $\tau_2 = t$ is non-zero, then $\epsilon_{ij} \p_i a_{2j}$ must be even under time reversal. Thus we  choose the action of time reversal on the $a^I_i$ to be 
$a^I_i\to T_{IJ}a^J_i$ with 
\be
 T=\lp\begin{array}{cc}
                                                        -1 & 0 \\
                                                        0  & 1 \end{array}\rp
\ee
As described in Ref. \onlinecite{Levin} we also need to describe the transformation of the quasiparticle creation operators. This is conveniently accomplished by using the standard edge theory that corresponds to the bulk Chern-Simons Lagrangian: 
\be
{\cal L}   =  \frac{1}{4\pi}\left(K_{IJ} \p_x \phi_I \p_t \phi_J +.......\right) +  \frac{1}{2\pi} \epsilon_{\mu\nu} \tau_I \p_\mu \phi_I A_\nu
\ee
Quasiparticle creation operators corresponding to $l = (1, 0)$ and $l = (0,1)$ are $e^{i\phi_1}$ and $e^{i\phi_2}$ respectively. The time reversal transformation of $a_{Ii}$ fixes the transformation of $\phi_I$ upto an overall phase. Thus we write 
\begin{eqnarray}
e^{i\phi_1} &  \rightarrow &  e^{i\left(\phi_1 +  \alpha_1\right)} \\
e^{i\phi_2} & \rightarrow & e^{-i\left(\phi_2 + \alpha_2\right)}
\end{eqnarray}
However by a shift of $\phi_1$ we can always set $\alpha_1 = 0$. This is not possible for $\alpha_2$. A further constraint comes from requiring that all physical operators transform such that ${\cal T}^2 = 1$. In particular ${\cal T}^2$  should take $e^{2i\phi_2} \rightarrow e^{2i\phi_2}$. This imposes the restriction that $\alpha_2 = \frac{\pi x}{2}$ with $x = 0, 1$. 
If $x = 0$ then the particle created by $e^{i\phi_2}$ is a Kramers singlet. If $x = 1$ however ${\cal T}^2$ takes $e^{i\phi_2} \rightarrow - e^{i\phi_2}$ so that the particle is a Kramers doublet. 

Thus within this $2 \times 2$ K-matrix we have four possible states corresponding to the four possible values of the pair $t, x$. 
In terms of Table \ref{u1rtz2t} these correspond to the four phases $e0m0, eT, eC, eCmT$.  Actually a fifth phase $eCT$ is also allowed in strict 2d but requires a $4 \times 4$ K-matrix. To see why this is so it is useful to better understand the physics of the 4 states described so far. 

First note that with the $K$-matrix in Eqn. \ref{z2k} the edge phase fields $\phi_{1}, \phi_2$ satisfy commutation relations such that the fields $f_{\pm} = e^{i\left(\phi_1 \pm \phi_2 \right)}$ satisfy fermion anti commutation relations. Indeed these correspond to $l = (1, 1), l = (1, -1)$ and describe the bulk fermionic $\epsilon$ particle. $f_{\pm}$ are the right and left moving fermions of the one dimensional edge Luttinger liquid theory.  Under a global $U(1)$ symmetry rotation $U_\theta$  by angle $\theta$ and time reversal, the $f_\pm$ transform as 
\begin{eqnarray}
U_\theta^\dagger f_\pm U_\theta &  = &   e^{i\frac{t\theta}{2}} f_\pm \\
{\cal T}^{-1} f_\pm {\cal T} & = & e^{\mp i\frac{\pi x}{2}} f_\mp
\end{eqnarray}

Note that as the $\epsilon$ particle may be regarded as a bound state of $e$ and $m$ , it has quantum numbers 
$\sigma_\epsilon = \sigma_e \sigma_f$ and $\mu_f = \mu_e \mu_f$. For the four cases described above in terms of edge Lagrangians this is  consistent with the symmetry transformation of the edge fermion fields.

Further insight is obtained by understanding how the four phases corresponding to the four choices of $(t,x)$ are obtained within a slave particle (parton) construction in the bulk. 
Consider  a slave particle (parton) construction obtained by writing the boson operator $b_r = a_r s_r$ at each site $r$ of the lattice. Here $a_r$ destroys a bosonic parton with charge $1$ while $s_r$ is an Ising parton (with charge $0$). We may take them to belong to the $e$ sector. Under time reversal $a_r, s_r$ remain invariant. Further as the $a_r, s_r$ carry only integer charge, $\sigma_e = \mu_e = 1$.  First we take the $a_r$ to form a simple bosonic Mott insulator and $s_r$ to form a  simple Ising paramagnet. Then the vison 
of the $Z_2$ gauge field associated with the slave particle construction will have trivial quantum numbers so that $\sigma_m = \mu_m = 1$. This corresponds to Phase $e0m0$ in Table. \ref{u1rtz2t}. 

Phase $eT$ can be likewise constructed if we start with two species of physical bosons $b_{1,2}$ and require $b_1\leftrightarrow b_2$ under time-reversal (e.g. spin-half bosons). Then we write the boson operators as $b_{1,2}=a_{1,2}s_{1,2}$ and put the system into a state such that time reversal is implemented through $(a_1,a_2)\to(-a_2,a_1)$ and $(s_1,s_2)\to(-s_2,s_1)$. The $e$ particles in this phase ($a_{1,2}$ and $s_{1,2}$) are Kramer's doublets, while the $m$ particle (the vision) transforms trivially under time-reversal. Nothing carries fractional charge in this phase.

Phase $eC$ is a familiar one and can be obtained in a slave particle construction by writing $b_r = \Phi_r^2$. The $\Phi_r$ destroys a charge-$1/2$ bosonic parton (denoted ``chargon" in the literature). Under time reversal $\Phi_r$ is invariant. Explicit microscopic models for the corresponding phase were studied in Refs. \onlinecite{bosfrc,bosfrc3d,SenthilFisher} with the standard implementation of time reversal symmetry for bosons ($b \rightarrow b$). 

Phase $eCT$ is also a familiar one. It can be obtained through a parton construction by writing the boson operator as $b_r = \epsilon_{\alpha\beta} f_{r\alpha} f_{r\beta}$ with $f_{r\alpha}$ a fermion. We will refer to $\alpha = 1,2$ as a pseudospin index.   The fermions carry charge -$1/2$. Time reversal is implemented through $f_{r\alpha} \rightarrow i\left(\sigma_y\right)_{\alpha\beta} f_{r\beta}$. Now consider a mean field ansatz where the fermion $f_{r\alpha}$ forms a (topologically trivial) band insulator that preserves time reversal but does not conserve any component of the fermion pseudospin.  The result is a $Z_2$ topologically ordered state with symmetry implemented as defined for Phase 4. 

Phase $eCmT$ is obtained from the same parton construction as for Phase $eCT$ but when the $f_{r\alpha}$ band structure is topologically non-trivial, {\em i.e} the fermions form a  2d topological insulator. Then a $\pi$ flux seen by the fermions (which we may take to be the m particle) is known to bind a Kramers doublet\cite{ranetal}. Indeed in the edge theory above if we 
choose $t = 1, x = 1$ the edge Lagrangian becomes identical to that of a fermionic topological insulator formed by the $\epsilon$ particle. 
Thus this parton construction has the symmetries of Phase 5. Three dimensional analogs of these phases were studied in Refs. \onlinecite{swingle3dfti,levin3dfti}.

It is clear now that the phase $eCT$ can exist in strictly $2d$ systems but is not captured by a Chern-Simons/edge theory description with a $2 \times  2$ K-matrix. This can also be seen by noting that since the physical charge is invariant under time-reversal, one cannot have a particle that's non-trivial under both $U(1)$ and ${\cal T}$ symmetries within this $2\times2$ $K$-matrix formulation.  

Note that for $eCmT$ the edge theory is gapless so long as the global symmetry is preserved. In contrast for the phases $e0m0, eT, eC$ the edge theory can be gapped by adding symmetry allowed perturbations. Similarly from the parton construction we know that though the $\epsilon$ particle carries the same quantum numbers for both $eCmT$ and for $eCT$ the edge theory for $eCT$ can be gapped. From the theory of the fermion topological insulator it follows that trivial band structure for the $\epsilon$ can be built up from the topological band structure by taking 2 copies and allowing all symmetry allowed perturbations. This suggests that the the minimal description of $eCT$ uses a $4\times 4$ 
$K$-matrix. Specifically consider 
\be
K=\lp\begin{array}{cccc}
          0 & 0 & 1 & 1 \\
          0 & 0 & 1 & -1 \\
          1 & 1 & 0 & 0 \\
          1 & -1 & 0 & 0 \end{array}\rp, T=\lp\begin{array}{cccc}
                                                                        -1 & 0 & 0 & 0 \\
                                                                        0 & 1 & 0 & 0 \\
                                                                        0 & 0 & 0 & 1 \\
                                                                        0 & 0 & 1 & 0 \end{array}\rp
\ee
with the charge vector $\tau=(0,1,0,0)$. Time reversal is implemented on the edge boson fields $\phi_I$ through $\phi_I \rightarrow T_{IJ} \left(\phi_I + \alpha_I \right)$ with $\alpha=(0,0,\pi,0)$. It is readily seen that this describes the $eCT$ phase. 

In passing we note that we can easily generate other $2d$ $Z_2$ topological phases with this symmetry by simply adding a layer of the $2d$ SPT phase allowed with $U(1) \rtimes Z_2^T$ symmetry to one of the 5 examples discussed above.  This obvious extension does not affect our subsequent discussion and we will not consider it further. 

In Appendix \ref{2dkz2} we explain in detail why all the other phases are not possible within the $K$-matrix formulation. In the next section we argue that, independent of the $K$-matrix formulation, the existence of those phases on SPT surfaces implies their non-existence in strict $2d$ systems.

For now we make a few comments. Note that in the first four phases  the m particle has trivial quantum numbers. It is only natural  that such states where one of the e or m particles have trivial quantum numbers have trivial quantum numbers can always be realized in strictly 2d systems. From such states we can always destroy the $Z_2$ topological order by condensing the m particle to produce a trivial symmetry preserving insulator. This will not be possible for states that can only be realized at the surface of 3d SPT phases. In Phase $eCmT$ both the e and m carry non-trivial quantum numbers. Despite this as we have seen it can be realized in strict 2d. 

Now lets move to the last 5 phases of Table. \ref{u1rtz2t}.  Ref. \onlinecite{avts12} showed that phases $eTmT$ and $eCmC$ both arise at the surface of 3d SPT phases. To discuss the other phases we first define the concept of ``surface equivalence" of topologically ordered phases. 

\subsection{Surface Equivalence}
We say that two topologically ordered states at the surface of a 3d SPT phase are ``surface equivalent" if one can be obtained from the other by combining with a strictly 2d states with the same symmetry. The notion of combining two states will be described in detail below.  Consider two $Z_2$ topologically ordered states - say states A and B - with distinct realizations of the global symmetry. This means that at least one of the $e,m$ particles transform differently under the global symmetry for the two states. Assume now that A and B have the same symmetry for the e particle or - in obvious notation - that $(\sigma_{eA}, \mu_{eA}) = (\sigma_{eB}, \mu_{eB}) \equiv (\sigma_e, \mu_e)$. Then  we must have 
$(\sigma_{mA}, \mu_{mA}) \neq (\sigma_{mB}, \mu_{mB})$. 

Now consider the composite system $A + B$. We allow A and B to couple through all symmetry allowed short ranged interactions. For weak interaction strengths the 2 states will be decoupled, and the combined system will have deconfined $Z_2 \times Z_2$ topological order.  However for stronger interactions $e_A$ can mix with $e_B$ as they have the same symmetry. This partially confines the $Z_2 \times Z_2$ topological order to a simpler topological ordered state with just a single deconfined  $Z_2$ gauge structure. We will denote this new phase $A \oplus B$.  In this new state the $m$ particles of A and B will be confined together to produce a new particle $m_{A \oplus B} \sim m_A m_B$. Thus  $A \oplus B$ has the quantum numbers $(\sigma_e, \sigma_{mA}\sigma_{mB}, \mu_e, \mu_{mA}\mu_{mB})$. 

This concept of combining phases enables us to see several equivalences in Table \ref{u1rtz2t}. For instance it is clear that Phase $eCT$ can be obtained as $eC \oplus eT$ (by letting the m particles mix).  Let us now consider surface equivalence. Phase $eCmC$ and $eCmT$ share the same quantum numbers for the $e$ particle. Thus we may combine them to produce a new $Z_2$ phase $ eCmC \oplus eCmT$ which, by inspection, has the same symmetries as Phase $eCTmC$ (after a relabeling of $e$ and $m$). This means that Phases $eCmC$ and $eCTmC$ are surface equivalent. Specifically consider the 3d SPT phase with Phase $eCmC$ as its surface topological ordered state. We may then deposit a layer of Phase $eCmT$ (which is allowed in strict 2d) at its surface, and then let the $e$ particles mix. This mixing will induce a surface phase transition where the surface topological order becomes that of Phase $eCTmC$. It follows that  Phase $eCTmC$ can also only be realized at the surface of  the 3d SPT boson insulator. 

Similarly the $m$ particle of Phase $eTmT$ has the same quantum numbers as the $m$ particle of Phase $eCmT$. Letting them mix we get Phase $eCTmT$. Phase $eCTmCT$ is also readily seen to be $eCTmC \oplus eCTmT$. 

Thus we see that the last 5 phases of Table \ref{u1rtz2t}  are all obtained at the surface of 3d SPT phases. All these 5 phases are obtained from two ``root" phases (Phase $eTmT$ and $eCmC$) by combining with phases that are allowed in strict 2d or with each other.

It is interesting to notice that the realization of the 5 phases at the SPT surfaces implies their absence in strict $2d$ systems, independent of $K$-matrix consideration. One can understand this as follows: if a surface state can also be realized in strict $2d$, then one can deposit such a $2d$ system onto the surface. The quasi-particles in the two systems (call them $(e_1,e_2)$ and $(m_1,m_2)$) will then have exactly the same symmetry properties, and the bound states of two particles of the same kind in the two systems ($e_1e_2$ and $m_1m_2$) will be trivial under all symmetries. Moreover, $e_1e_2$ and $m_1m_2$ are mutual bosons to each other. Hence one can condense both $e_1e_2$ and $m_1m_2$ without breaking any symmetry. However, this will confine all the fractional quasi-particles since any one of them will have mutual $\pi$-statistics with either $e_1e_2$ or $m_1m_2$, and the surface will become a trivial phase, i.e. symmetric, gapped and confined. By definition, the corresponding bulk cannot be a SPT state. Hence the states at SPT surfaces must not be realizable in strict $2d$. This will have interesting implications for $2d$ systems, and an example will be given toward the end of this paper.


It is also interesting to view this result from a different  point of view which inverts the logic followed above. Consider the problem of identifying 3d boson SPT states with this symmetry.  The results of this section show that there are precisely two distinct  `root'  $Z_2$ topological orders that can only occur at the surface of SPT phases. phases . This then gives us two ``root" 3d SPT states with this symmetry. This is the same conclusion arrived at by direct consideration of surface theories in Ref. \onlinecite{avts12}, and ties in nicely wth the formal cohomology classification (which also gives 2 root states).  Note in particular that of the 2 root states $eTmT$ is simply inherited from the 3d SPT with $Z_2^T$ symmetry alone. Thus the only non-trivial SPT state that is unique to the extra $U(1)$ symmetry is the one with surface topological order $eCmC$ as was suggested in Ref. \onlinecite{avts12}. 

\section{Topological $Z_2$ spin liquids}
\label{z2t}
Here we repeat the excercise above for symmetries appropriate to quantum spin systems. We consider two cases: symmetry $U(1) \times Z_2^T$ and symmetry $Z_2^T$. The former describes time reversal symmetric quantum spin Hamiltonians with a conserved component of spin. In the latter we only assume time reversal symmetry.  The consistent symmetry assignments for $Z_2$ topological order with bosonic $e$ and $m$ particles is given in Tables. \ref{u1tz2t} and \ref{z2t}. 

Let's first consider $U(1)\times Z_2^T$, in which case the $U(1)$ charge goes to minus itself under time-reversal. The analysis of Ref. \onlinecite{hermele} again gives the same 10 phases as before and we will use the same labels. However a difference appears in the K-matrix classification. For this symmetry class we will see that a $2\times 2$ $K$-matrix is enough to describe all the $2d$ states. We have again
\be
K=\lp\begin{array}{cc}
          0 & 2 \\
          2 & 0 \end{array}\rp, T=\lp\begin{array}{cc}
                                                        -1 & 0 \\
                                                        0  & 1 \end{array}\rp
\ee
but now the charge vector must be taken to be $\tau = (t, 0)$ with $t = 0, 1$ due to the different transformation of the density of the global $U(1)$ charge under time reversal. 
Time reversal on the edge boson fields continues to be  implemented as $\Phi_I \rightarrow T_{IJ}\left(\phi_I + \alpha_I \right)$ with $\alpha = (0, \frac{\pi x}{2})$ ($x = 0, 1$). 
With this symmetry implementation we see that the edge field $e^{i\phi_2}$ creates a particle that can either carry $1/2$ charge or be a Kramers doublet or both. The other edge field $e^{i\phi_1}$ creates a particle with trivial quantum numbers. This leads to four phases corresponding to $e0m0,eT,eC$ and $eCT$. 

The standard slave boson/fermion construction of $Z_2$ spin liquids - as in the classic work of Refs. \onlinecite{ReSaSpN,wen91} - give (when the spin symmetry is $U(1)$) the state $eCT$. The spinon in these constructions both carries spin-$1/2$ and is a Kramers doublet. The easy axis Kagome lattice spin model of Ref. \onlinecite{bfg} provides an explicit microscopic model for a $Z_2$ spin liquid with $U(1) \times Z_2^T$ symmetry. In the standard interpretation the spin $S_z$ of that model labels the two members of a Kramers doublet states of a microscopic Ising spin. Time reversal is then implemented in terms of the spin operators as usual through $\vec S \rightarrow -\vec S$. In that case in the $Z_2$ spin liquid phase the spinons are readily seen to both carry $S_z = \pm \frac{1}{2}$ and be Kramers doublets to realize the $eCT$ class. There is a different implementation of time reversal symmetry in this easy axis Kagome spin model.  If the $S_z$ labels two members of a microscopic non-Kramers doublet then we must interpret the $\vec S$ as a `pseudo spin' $1/2$ operator that acts in this two dimensional Hilbert space at each site. Time reversal takes $S_z \rightarrow - S_z, S^+ \rightarrow S^-$. In that case the spinons in the $Z_2$ spin liquid phase will have spin $S_z = \pm \frac{1}{2}$ but will be Kramers singlets. Thus we have a realization of the $eC$ phase in the model of Ref. \onlinecite{bfg}.

Phase $eCmT$ which was allowed earlier in Sec. \ref{bu1rtz2} now does not appear. Physically this is because the "topological band" in the $U(1)\rtimes Z_2^T$ symmetry becomes trivial in the $U(1)\times Z_2^T$ case. The easiest way to see this is to consider the edge theory, which has two counter-propagating fermions. With $U(1)\times Z_2^T$ symmetry, one can mix the two fermions (hence gap out the edge) without breaking any symmetry, even if the fermions form Kramer's pairs.

 The absence of the $eCmT$ phase in strict $2d$ modifies the equivalence relation established in last section. In particular, the last three phases in Table \ref{u1rtz2t} will not be equivalent to either $eTmT$ or $eCmC$. Actually in Ref. \onlinecite{avts12}, three distinct `root' SPT phases were discussed  corresponding to those with surface topological orders$eTmT$, $eCTmT$ and $eCTmCT$.  The last three phases in Table \ref{u1tz2t} are thus surface topological orders corresponding to other SPT phases that may be obtained by combining these root phases. This is in perfect agreement with the results of Ref. \onlinecite{avts12}.

 \begin{table}[htdp]
\begin{tabular}{|c|c|c|c|c|c|}
\hline
Phase & $\sigma_e$ & $\sigma_m$ & $\mu_e$ &$ \mu_m$ & Comments \\ \hline
$e0m0$ & $1$ & $1$ &  $1$ & $1$ & No fractionalization \\ \hline
$eT$ & $ 1$ & $1$ & $-1$ & $1$ & No fractional charge but Kramers \\ \hline
$eC$ &  $-1$ & $1$ &  $1$ & $1$ & $b = \Phi^2$ \\ \hline
$eCT$ &  $-1$ & $1$ & $-1$ &  $1$ & $b = \epsilon_{\alpha\beta} f_{\alpha} f_{\beta}$ \\  \hline
\hline
$eTmT$ & $1$ & $1$ &  $-1$ & $-1$ & 3d SPT surface \\ \hline
$eCTmT$ & $ -1$ & $1$ & $-1$ & $-1$ & 3d SPT surface \\ \hline
$eCTmCT$ &  $-1$ & $-1$ &  $-1$ & $-1$ &   3d SPT surface \\ \hline
$eCmT$ & $ -1$ & $1$ & $1$ & $-1$ & $eTmT \oplus eCTmT$ \\ \hline
$eCTmC$ & $ -1$ & $-1$ & $-1$ & $1$ &   $ eCTmT \oplus eCTmCT$  \\  \hline 
$eCmC$ &  $-1$ & $-1$ & $1$ &  $1$ &   $ eCTmC \oplus eCmT$  \\
\hline
\end{tabular}
\caption{ Symmetry action ($U(1) \times Z_2^T$) for  $Z_2$ topological ordered states. The first 4 are allowed in strict 2d while the last 6 can only be realized at surface of 3d SPT phases (or derived from them)}
\label{u1tz2t}
\end{table}%

Next we consider $Z_2^T$ symmetry alone, which is much simpler. It is straightforward to see that the phases $e0m0$ and $eTm0$ can be realized in strict $2d$, while $eTmT$ can only appear on an SPT surface. The corresponding table is simply a subset of the previous two. 

 \begin{table}[htdp]
\begin{tabular}{|c|c|c|c|}
\hline
Phase &  $\mu_e$ &$ \mu_m$ & Comments \\ \hline
$e0m0$ & $1$ & $1$ & No fractionalization \\ \hline
$eT$ &  $-1$ & $1$ & Kramers \\ \hline
\hline
$eTmT$ &  $-1$ & $-1$ & 3d SPT surface \\ 
\hline
\end{tabular}
\caption{ Symmetry action ($Z_2^T$) for  $Z_2$ topological ordered states. The first two are allowed in strict 2d while the last one can only be realized at surface of 3d SPT phases}
\label{z2t}
\end{table}%

\section{All fermion $Z_2$ liquids}
\label{beyondcoh}
 
We now extend our analysis to a very interesting  topological order where there are  three distinct topological quasi-particles, all of which are fermions $f_{1,2,3}$, and there's a mutual $\pi$-statistics between any two of them. This can be viewed as a variant of the usual $Z_2$ liquid, in which both the $e$ and $m$ particles become fermions. Since they have a mutual $\pi$-statistics, the bound state $\epsilon=em$ is still a fermion and has $\pi$-statistics with both $e$ and $m$. 

The statistics of this phase is perfectly compatible with time-reversal symmetry, but the realization in strict $2d$ turns out to be always chiral and hence breaks time-reversal. One way to understand this is to start from a conventional $Z_2$ topologically ordered liquid with bosonic $e$ and $m$ particles. Then put the fermionic $\epsilon$ particle  into a band structure such that the vison also becomes a fermion. This may be fruitfully discussed in terms of the edge Lagrangian for  the $\epsilon$ field. The vison operator appears as a `twist'  field that creates a $\pi$ phase shift for $\epsilon$.  For a single branch of chiral (complex) fermion on the edge $e^{i\phi_{L,R}}$ the twist operator is $e^{i\phi_{L,R}/2}$. This has  conformal spin $\pm1/8$ so that in this case the vision is an anyon with fractional statistics. Very generally take a theory with $n_R$ right moving and $n_L$ left moving fermions all of which correspond to the same bulk $\epsilon$ particle which see a single common vison. This acts as a common twist field for all the edge fermions and hence has conformal spin $\frac{n_R - n_L}{8}$. Therefore to make the vison fermionic one needs $n_L-n_R=4$  mod $8$.  One such realization is given by the $4\times 4$ $K$-matrix
\be
K=\lp\begin{array}{cccc}
          2 & -1 & -1 & -1 \\
          -1 & 2 & 0 & 0 \\
          -1 & 0 & 2 & 0 \\
          -1 & 0 & 0 & 2 \end{array}\rp
\ee
which has chiral central charge $4$. Since the chiral central charge is non-zero this phase clearly cannot arise in a time reversal invariant strictly 2d system. 

However Ref. \onlinecite{avts12} suggested that such an all fermion $Z_2$ topological order can arise at the surface of a 3d SPT phase with time reversal symmetry. 
In this state if the surface is gapped by breaking time reversal symmetry  then there is a quantized thermal Hall conductivity $\kappa_{xy} = \pm 4$. However if time reversal symmetry is present and the surface is gapped, there will be surface topological order. Ref. \onlinecite{avts12} proposed that this is a  $Z_2$ topological order which is a time reversal symmetric all fermion state. To understand why this is reasonable consider starting from the all fermion surface topological ordered state. What should we do to confine all the fermion excitations in the surface? It is clear from the discussion above that if we take one of the fermions and put it in a Chern band such that the surface $\kappa_{xy} = \pm 4$ then the other two topological quasiparticles will become bosons. These bosons can now be condensed to get a confined surface state. However this clearly requires broken time reversal symmetry and will give a $\kappa_{xy} = \pm 4$ which is indeed the right $Z_2^T$ broken surface state for this proposed SPT.  
This kind of 3+1-D SPT phase with $Z_2^T$ symmetry is not present in the cohomology table of Ref. \onlinecite{chencoho2011}.   
Including this 3d SPT surface state and using the language in the last few sections, we have a new table (Table \ref{fz2t}). 

 \begin{table}[htdp]
\begin{tabular}{|c|c|c|c|}
\hline
Phase &  $\mu_e$ &$ \mu_m$ & Comments \\ \hline
$e^f0m^f0$ & $1$ & $1$ & All fermions, singlets \\ \hline
$e^fTm^fT$ &  $-1$ & $-1$ & $e^f0m^f0\oplus eTmT$ \\ \hline
\end{tabular}
\caption{ Symmetry action ($Z_2^T$) for all- fermionic $Z_2$ states. Both states can only be realized at surfaces of 3d SPT phases}
\label{fz2t}
\end{table}%

The second phase in the table is obtained from the first by adding a usual $Z_2$ liquid in the $eTmT$ phase, then condense the bound state of the $\epsilon^f=e^fm^f$ in the fermionic liquid and the $\epsilon=em$ in the $eTmT$ liquid. Since the $eTmT$ phase cannot be realized in strict $2d$, the two phases $e^f0m^f0$ and $e^fTm^fT$ should be viewed as inequivalent, hence give rise to two distinct  SPT phases with time-reversal symmetry in addition to the one with the $eTmT$ surface $Z_2$ topological order. Thus in total with $Z_2^T$ symmetry we actually have 3 non-trivial SPT phases corresponding to a $Z_2^2$ classification. (The cohomology classification of Ref. \onlinecite{chencoho2011} gives instead a $Z_2$ classification). 

\section{Constructing SPT with coupled layers of $Z_2$ liquids}
\label{clayer}
From the considerations in the previous sections, it is clear that to construct a 3+1-D SPT state, we only need to construct the corresponding topological order on the surface but have a confined bulk with gapped excitations. In this section we give one such 
explicit construction using coupled layers of $2d$ $Z_2$ liquids. Specifically we consider a system of stacked layers where each layer realizes a $Z_2$ topological order that is allowed in strictly 2d systems. Then we couple the different layers together in such a way that the bulk is confined and gapped. But we show that the surface layer is left unconfined and further 
corresponds to the surface $Z_2$ topological order of an SPT phase. A similar coupled-layer construction of the free fermion topological insulator was proposed\cite{layerti} to obtain the sigle Dirac cone on the surface. We first illustrate this by constructing the $eTmT$ with $Z_2^T$ symmetry, and it will be clear later that this can be generalized to all the SPT states mentioned in this paper. 

\begin{figure}
 \includegraphics[scale=0.5]{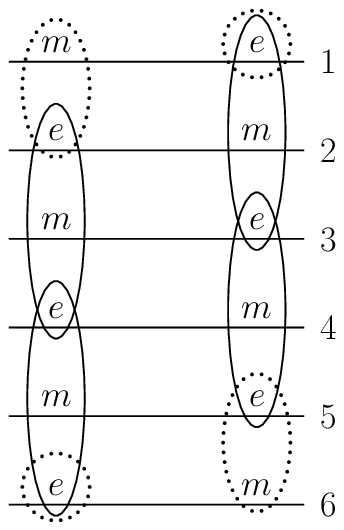}
\caption{Coupled-layer construction of SPT states. The particle composite in the ellipses are condensed, and only the four surface particles 
in the dotted ellipses survived as deconfined topological quasi-particles.}
\label{layer}
\end{figure}

Consider stacking $N$ layers of $Z_2$ liquids in the $eT$ state which is allowed in strictly $2d$.  Now  turn on an inter-layer coupling to make the composite particles $e_im_{i+1}e_{i+2}$ condensed, where $i$ is the layer index running from 
$1$ to $N-2$.  Note that the $e_im_{i+1}e_{i+2}$ all have bosonic self and bosonic mutual statistics so that they may be simultaneously condensed. As illustrated in Figure \ref{layer}, this procedure confines all the non-trivial quasi-particles in the bulk. However four 
particles on the surfaces survive as the only deconfined objects: $e_1,m_1e_2,e_N,m_{N}e_{N-1}$. Notice that $e_1$ and $m_1e_2$ are mutual semions and have self-boson statistics. Thus they form a $Z_2$ liquid at the top surface. Similarly  $e_N$ and $m_Ne_{N-1}$ are self bosons, have mutual semion statistics and form a $Z_2$ liquid at the bottom surface. The key point however is that all these particles have ${\cal T}^2=-1$. Thus either surface is in the 
$eTmT$ state though the bulk has no exotic excitations.  By the analysis above we identify this with the 3d  SPT state with $Z_2^T$ symmetry. 

This construction can be immediately generalized to other SPT states. For example, to get the $eCmC$ (or $eCTmCT$) state with $U(1)\rtimes Z_2^T$ or $U(1)\times Z_2^T$ symmetry, 
just stack layers of $eC$ (or $eCT$) states and condense $e_im_{i+1}e_{i+2}$. 

Most interestingly, the all-fermion $Z_2$ surface topological state with global $Z_2^T$ symmetry, which is quite hard to construct using other methods, 
can also be constructed in this way: simply start with stacked 2d $Z_2$ liquids where all particles $e, m, \epsilon$ are invariant under $T$-reversal. Such a $Z_2$ state is obviously allowed in strict $2d$. Now condense $\epsilon_im_{i+1}\epsilon_{i+2}$ instead of $e_im_{i+1}e_{i+2}$ in the above constructions, where $\epsilon_i=e_im_i$ is the 
fermion in the $2d$ $Z_2$ gauge theory. Again the $\epsilon_i m_{i+1} \epsilon_{i+2}$ have both self and mutual boson statistics so that they may be simultaneously condensed. This confines all bulk topological quasiparticles. The surviving surface quasi-particles will be $\epsilon_1,m_1\epsilon_2$ at the top surface and $\epsilon_N,m_N\epsilon_{N-1}$ at the bottom surface. These particles  are all fermions, and the two particles at either surface have mutual semion statistics. It follows that either surface realizes the all-fermion $Z_2$ topological order but now in the presence of $Z_2^T$ symmetry.  We have thus explicitly constructed the SPT phase discussed in Section. \ref{beyondcoh}. 

This coupled layer construction gives very strong support to the results of Ref. \onlinecite{avts12} on the various SPT phases.  In particular it removes any concerns on the legitimacy of the state of Sec. \ref{beyondcoh} with $Z_2^T$ symmetry not currently present in the cohomology classification. 

\section{Relation with bulk topological field theories}
\label{rbtft}
Here we provide an understanding of the results obtained above from topological field theories in the bulk. It was shown in Ref. \onlinecite{avts12} that  bosonic topological insulators in $3d$ with $\left (U(1)\right)^N$ symmetry has a bulk response to external `probe' gauge fields  $A^I$ characterized by a $\theta$-term with $\theta=\pi$:
\be
L_{\theta}=\frac{\theta}{8\pi^2}K_{IJ}\epsilon^{ijkl}\p_i A^I_j\p_k A^J_l.
\ee
If under symmetry transformations (e.g. time-reversal) the $\theta$-angle transforms as $\theta\to-\theta$, then the $\theta=\pi$ term is symmetric in the bulk, but on the boundary it reduces to a (mutual) Chern-Simons term with symmetry-violating responses. This was a familiar issue in the non-interacting fermionic topological insulator, where a single Dirac cone was introduced on the boundary to cancel the time-reversal violating response through parity anomaly.

In our cases let us understand how this works out when the surface is in a symmetry preserving gapped topological ordered phase of the kind studied in this paper.  We will show that  the symmetries of the topological order on the boundary are such as  to cancel the Chern-Simons response arising from the $\theta$-term. To illustrate the idea we take $K_{IJ}=\sigma_x$ which applies to a large class of SPT phases in $3d$. This will give a mutual Chern-Simons term on the boundary
\be
L_{CS}=\frac{1}{4\pi}\epsilon^{ijk}A_{1,i}\p_jA_{2,k}.
\ee
This term alone would give a response that breaks time-reversal symmetry. To cure it we put a $Z_2$ topological liquid on the boundary, with the $e$ and $m$ particles coupling to $A_{1,2}$ respectively. The Lagrangian is given by
\be
L_{Z_2}=\frac{1}{\pi}\epsilon^{ijk}a_{1,i}\p_ja_{2,k}+\frac{1}{2\pi}\lp\epsilon^{ijk}A_{1,i}\p_ja_{1,k}+\epsilon^{ijk}A_{2,i}\p_ja_{2,k}\rp.
\ee
Integrating out $a_{1,2}$ induces a mutual Chern-Simons term for $A_{1,2}$ which exactly cancels what arose from the bulk $\theta$-term and  hence restores time-reversal symmetry.

The topological ordered states with symmetry that are forbidden in strict $2d$ realize the symmetry in an `anamolous' way. The corresponding topological field theories cannot be 
given consistent lattice regularizations which implement the symmetry in a local manner. The discussion in this section illustrates how these theories can nevertheless be given a higher dimensional regularization as the boundary of a non-anomalous field theory. 
This has the same essence with the anomaly cancellation in free fermion TI.  It will be interesting for the future to have criteria to directly identify such `anamolous' symmetry in a topological field theory. 

\section{Constraints on gapless 2d quantum spin liquids: Absence of Algebraic Vortex Liquids}
\label{avl}
We now turn to gapless quantum liquids in two space dimensions. Examples are gapless quantum spin liquid phases of frustrated quantum magnets, or non-fermi liquid phases of itinerant fermions or bosons. Symmetry plays a very crucial role in the stability of these phases. The example of the topologically ordered gapped states considered in previous sections lead us to pose the question of what kinds of putative gapless phases/critical points are allowed to exist with a certain symmetry in strictly 2d systems.  
First of all we note that in contrast with gapped topologically ordered phases global symmetries typically play a much more important role in protecting the gaplessness of a phase. 
The symmetry may forbid a relevant perturbation to the low energy renormalization group fixed point that, if present, may lead to a flow to a gapped fixed point. Here however we are interested in a more general question. We wish to consider gapless fixed points that can be obtained by tuning any finite number of relevant perturbations. This includes not just bulk 2d phases but also critical or even finitely multi critical quantum systems. We are particularly interested in such gapless 2d fixed points with symmetry  that cannot exist in strict 2d but may  only exist at the surface of a 3d insulator (SPT or otherwise).  

To set the stage consider a simple and familiar example in a  free fermion system. The surface of the celebrated time reversal symmetric electron topological insulator (symmetry $U(1) \rtimes Z_2^T$ has an odd number of Dirac cones. Such a gapless state cannot exist in strict 2d fermion systems with the same symmetry even as a multi critical point.  However if we give up time reversal symmetry this state is allowed as a critical point in strict 2d. An example is provided by a 2d free fermion model poised right at the integer quantum Hall transition. Thus symmetry provides a strong restriction on what gapless fixed points are allowed in strict 2d. 

We focus now on a very interesting gapless state proposed\cite{avl} to exist in strict 2d in frustrated $XY$ quantum magnets (symmetry $U(1) \times Z_2^T$) or in boson systems (symmetry $U(1) \rtimes Z_2^T$). This state - dubbed an Algebraic Vortex Liquid (AVL) - was obtained in a dual vortex description by fermionizing the vortices and allowing them to be massless. A suggestive approximation was then used to derive a low energy effective field theory consisting of an even number of massless 2-component Dirac fermions (the vortices) coupled to a non-compact $U(1)$ gauge field. The AVL state has been proposed to describe quantum spin liquid states on the Kagome and triangular lattices. 
In terms of development of the theory of gapless spin liquids/non-fermi liquids the AVL proposal is extremely important. To date the only known theoretical route to accessing such exotic gapless phases of matter (in $d > 1$) is through a slave particle construction where the spin/electron operator is split into a product of other operators. If some of the resulting slave particles are fermions, they can be gapless. In contrast the AVL presents a new paradigm for a gapless highly entangled state which is likely beyond the standard slave particle approach. It is therefore crucial to explore and understand it thoroughly. 

We now argue that the AVL state cannot exist in strictly 2d models with either $U(1) \rtimes Z_2^T$  or $U(1) \times Z_2^T$ symmetry.  This is already hinted at by several observations. First it has never been clear how to implement time reversal in a consistent way in the AVL theory. The AVL is obtained from the usual bosonic dual vortex theory through a flux attachment procedure to fermionize the vortices. This leads to an additional  Chern-Simons gauge field that couples to the fermionized vortices. However this new gauge field can be absorbed into the usual dual gauge field to leave behind a three derivative term for a residual gauge field. It was argued that this three derivative  interaction is formally irrelevant in the low energy effective theory. This argument is delicate though. In the simplest context \cite{avl0} where such an approximation was made an alternate description\cite{tsfisher06} in terms of a sigma model revealed the presence of a topological $\theta$ term at $\theta = \pi$. The topological term also has three derivatives but its coefficient is protected by time reversal symmetry and does not flow under the RG. It's presence presumably crucially alters the physics of the model.  Thus one may worry about the legitimacy of the approximations
invoked to justify the AVL phase. 

Note that the fundamental issue that needs to be addressed with the AVL phase is whether it realizes symmetry in a manner that is allowed in 2d spin/boson systems.  This is of course the kind of question that is the essence of this paper. A final hint that the AVL phase may not exist in strict 2d comes from recent work\cite{avts12} showing that  gapless quantum vortex liquids with fermionic vortices can actually arise at the surface of time reversal symmetric 3D SPT phases. This strongly suggest that such phases cannot arise in strict 2d with the same symmetries. Below we will sharpen these arguments.

Consider the proposed effective field theory for the AVL phase: 
\begin{equation}
{\cal L} = \overline{\psi}_\alpha \left(i\slashed{\partial}_\mu - i\slashed{a}_\mu\right) \psi_\alpha + \frac{1}{2e^2}\left(\epsilon_{\mu\nu\lambda}\partial_\nu a_\lambda \right)^2 + \frac{1}{2\pi}a_\mu \epsilon_{\mu\nu\lambda}\partial_\nu A_\lambda^{ext}
\end{equation}
Here $\psi_\alpha$ ($\alpha = 1,.....2N$) are the fermionized vortices, $a_\mu$ is a non-compact $U(1)$ gauge field whose curl is $2\pi$ times the global $U(1)$ current, and $A_\mu^{ext}$ is an external probe $U(1)$ gauge field. Note that the vortices themselves do not carry physical $U(1)$ charge.  As mentioned above the realization of time reversal in terms of these fermionized vortex fields has always been a tricky issue for the AVL theory.  To sharpen the issue we now consider 
a phase that is accessed from the AVL phase by pairing and condensing the $\psi_\alpha$ fields.  In the original AVL literature\cite{avl} a number of phases proximate to the AVL were studied by assuming that four fermion interactions were strong enough to give a mass to the fermions. A number of different such mass terms leading to various symmetry breaking orders were examined. Here instead we imagine a mass term that corresponds to vortex pairing that preserves the global internal symmetry.

The vortex pair condensation will gap out the gauge field $a_\mu$ and will give an insulator. However as the $\psi_\alpha$ fields are vortices of the original boson this is a phase with $Z_2$ topological order.  The fermionized vortices survive as `unpaired' gapped quasiparticles in this topological phase. We may identify them with the $\epsilon$ particle which in this case has zero global $U(1)$ charge.  In the notation of previous sections $\sigma_\epsilon = 1$.  Furthermore the pair condensation will quantize the gauge flux $\vec \nabla \times \vec a$ in units of $\pi$ so that one of the topological quasiparticles (which we take to be the $e$ particle) has $1/2$ charge, {\em i.e} $\sigma_e = -1$.

Now it is clear from Tables \ref{u1rtz2t} and \ref{u1tz2t} of the previous sections that in strict 2d such a state can exist only if the $\epsilon$ particle also carries $1/2$ charge, {\it  i.e} $\sigma_\epsilon = -1$. However we just argued that the $Z_2$ topological ordered state realized from the AVL state has $\sigma_\epsilon = 1$, {\em i.e} it carries zero global $U(1)$ charge. It follows that such a $Z_2$ topological ordered state cannot exist in strict 2d. Note that we did not explicitly rely on time reversal symmetry in our analysis (though it is implicit in deciding which $Z_2$ states are allowed in 2d). 

Thus the $Z_2$ topological ordered states that descends from the AVL is not allowed to exist in strict $2d$. This then implies that the AVL itself cannot exist in strictly 2d systems so long as both global $U(1)$ and $Z_2^T$ symmetries are present. 

Can gapless quantum vortex liquids ever exist in strictly 2d? One option is to break time reversal symmetry. Then our arguments do not prohibit the formation of fermionic vortices which can then be in a gapless fluid state. Indeed such a gapless magnetic-field induced vortex metal state was proposed to exist in 2d superconducting films in Ref. \onlinecite{2dvrtxmtl}. A different option -  which we will elaborate elsewhere\cite{chongfrcvrtx} - that preserves internal symmetries is obtained by fractionalizing the vortices into fermionic partons which can then be gapless. 

\section{Time reversal symmetric $U(1)$ quantum  liquids in $3+1$ dimensions}
\label{3du1}
We now turn our attention to three dimensional highly entangled states with time reversal symmetry.   In three dimensions, interesting gapless quantum liquids with an emergent gapless $U(1)$ gauge field are possible\cite{wen01}. Explicit lattice models for such phases were constructed and their physics studied in Refs. \onlinecite{bosfrc3d,wen03,hfb04,3ddmr,lesikts05,kdybk,shannon}.  Interest in such phases has been revived following a recent  proposed realization\cite{balentsqspice}   in quantum spin ice materials on three dimensional pyrochlore lattices. It is thus timely to understand the possibilities for the realization of symmetry in such phases with emergent photons. Here we will restrict attention to time reversal symmetry in keeping with the theme of the rest of the paper. 

The excitation spectrum of the $U(1)$ quantum liquid consists, in addition to the gapless photon, point `electric' charges (the $e$ particle) and point `magnetic' charges (the $m$ particle or monopole).  We will only consider the situation in which both the $e$ and $m$ particles are gapped, and will focus on phases that can be realized in strictly $3d$ systems (as opposed to $U(1)$ phases allowed at the boundary of $4+1$ dimensional SPT phases). Following the discussion of previous sections, a simple restriction that ensures this is to assume that one of the $e$ or $m$ particles has trivial global quantum numbers and is a boson. Without  loss of generality we will assume that it is the $m$ particle. 

The low energy long wavelength physics of the $U(1)$ liquid state is described by Maxwell's equations. As usual they imply that the emergent electric and magnetic fields transform oppositely under time reversal. We will distinguish two cases depending on whether the electric field is even or odd under time reversal. 

\subsection{Even electric field}
\label{evene}

First we consider the case $\vec E \rightarrow \vec E, \vec B \rightarrow - \vec B$ under time reversal. This is what happens in the usual slave particle constructions of $U(1)$ spin liquids through Schwinger bosons or fermions. The electric field on a bond gets related to the bond energy which is clearly even under time reversal.  Consistent with this the magnetic field gets identified with the scalar spin chirality which is odd under time reversal.  Then the electric charge $q_e \rightarrow q_e$ and magnetic charge $q_m
\rightarrow - q_m$. Let us introduce creation operators $e^\dagger, m^\dagger$ for the $e$ and $m$ particles. With the assumption that $m^\dagger$ has trivial global quantum numbers and is a boson, it must transform under time reversal as 
\begin{equation}
{\cal T}^{-1}m^\dagger {\cal T} = e^{i\alpha_m} m
\end{equation}
However the phase $\alpha_m$ has no physical significance. It can be removed by combining ${\cal T}$ with a (dual) $U(1)$ gauge transformation that rotates the phase of $m$ 
(more detail follows in Sec. \ref{odde}). So we may simply set $\alpha_m = 0$. Let us consider now the $e$ particle.  If there is just a  single species of $e$ particle, then we must have 
\begin{equation}
{\cal T}^{-1} e^\dagger {\cal T} = e^{i\alpha_e} e^\dagger
\end{equation}
Now the phase $\alpha_e$ can be absorbed by redefining the $e$ operator and so we set $\alpha_e = 0$. The $e$ particle transforms trivially under $Z_2^T$. There are nevertheless two distinct phases depending on whether $e$ is a boson or a fermion. More phases are obtained by considering a $2$-component $e$ field: $e = (e_1, e_2)$.  The new non-trivial possibility is that this $2$-component $e$ field transforms as a Kramers doublet under $Z_2^T$: 
\begin{equation}
{\cal T}^{-1} e^\dagger {\cal T} = i\sigma_y e^\dagger
\end{equation}
Clearly we have ${\cal T}^{-2} e^{\dagger} {\cal T}^2 =- e^{\dagger}$ but the action of ${\cal T}^2$ on physical ({\em gauge invariant} local) operators gives 1. For instance $e^\dagger_1 e_2$ is a physical operator and we clearly have ${\cal T}^{-2} e^\dagger_1 e_2 {\cal T}^2 = e^\dagger_1 e_2$. When  $e$ is a Kramers doublet  it can again be either a boson or a fermion. The former is obtained in the standard Schwinger boson construction and the latter in the Schwinger fermion construction. Thus we have a  total of four possible phases corresponding to $e$ being a Kramers singlet/doublet with bose/fermi statistics and a boson monopole with trivial global quantum numbers. 

\subsection{$U(1)$ quantum liquids as monopole topological insulators}
The four $U(1)$ quantum liquids described above were distinguished by the symmetry and statistics of the $e$ particle. We now develop a very interesting alternate view point where 
we understand these four states as different SPT insulators of the bosonic monopole with trivial global quantum numbers. As the magnetic  charge is odd under time reversal, the monopole transforms under $U_g(1) \times Z_2^T$ where $U_g(1)$ is the gauge transformation generated by the monopole charge.   It is useful (though not necessary) to perform an electric-magnetic duality transformation: this exchanges the $e$ and $m$ labels: 
\begin{eqnarray}
e & \leftrightarrow & m_d \\
m & \leftrightarrow & e_d
\end{eqnarray}
We included a subscript $d$ on the right side to indicate that these are the dual labels. Now $e_d$ is a gapped boson that transforms under $U_g(1) \times Z_2^T$.  Thus we may regard the $U(1)$ quantum liquids as insulating phases of $e_d$ obtained by gauging the $U(1)$ part of a $U(1) \times Z_2^T$ symmetry.  Note that $e_d$ transforms under a linear 
({\em i.e} not projective) representation of $U_g(1) \times Z_2^T$. As discussed in previous sections such bosons can be in a number of different SPT phases.  We now study their fate when the $U(1)$ symmetry is gauged.

\subsubsection{Gauged bosonic SPT phases in $3d$}
\label{gb3d}
In $2d$ Ref. \onlinecite{levingu} studied the fate of  bosonic SPT insulators with discrete global unitary symmetry when that symmetry is gauged. It was shown that the result was a topologically ordered gapped quantum liquid with long range entanglement.  A general abstract discussion of such gauged SPT phases for unitary symmetry groups ({\em i.e} not involving time reversal) has also appeared\cite{janetwen}. Here we are interested in $3d$ SPT phases with $U(1) \times Z_2^T$ symmetry. A gauged $3d$ SPT phase with $U(1) \rtimes Z_2^T$ was also studied very recently in a beautiful paper\cite{mkf13}. Using the known $\theta = 2\pi$ electromagnetic response\cite{avts12}, it was argued that the monopole of this gauged SPT is a fermion, and this was used as a conceptual starting point to discuss the surface of this SPT.  Here we will discuss the gauged SPT from a different point of view that will enable us to also discuss SPT phases where the electromagnetic response has no $\theta$ term (necessary for the results in this subsection). 

Refs. \onlinecite{avts12,xuts13} show that a key distinction between different SPT phases with the same symmetry is exposed by considering the end points of vortex lines of the boson at the interface with the vacuum.  It will be convenient to label the SPT phases by their possible surface topological order (whether or not such order is actually present in any particular microscopic realization). For one simple example SPT phase (the one whose surface topological order is $eCmC$) these papers argued that a surface Landau-Ginzburg theory 
is obtained in a dual description in terms of fermionic vortices.  In another SPT phase (labeled by surface topological order $eCmT$) the surface vortex is a boson but is a Kramers doublet. By stacking  these two phases together we can get a third SPT phase where the surface vortex is a fermionic Kramers doublet. In contrast for topologically trivial insulators the surface vortex is a bosonic Kramers singlet. In Appendix \ref{duallgw} we describe these surface dual Landau-Ginzburg theories and their implied surface phase structure. We also provide an explicit derivation that is complementary to the arguments of Ref. \onlinecite{avts12}. 

Closely related to this we can also consider external point sources for vortex lines directly in the bulk. In the Hilbert space of the microscopic boson model the vortex lines do not have open ends in the bulk. So these external sources for vortex lines must be thought of as `probes' that locally modify the Hilbert space. These will behave similarly to the surface end points of vortices.  For example in the SPT labelled $eCmT$ Ref. \onlinecite{xuts13} shows that the ground state wave function is a loop gas of vortices where each vortex core is described as a Haldane spin chain. An externally imposed open end for a vortex string will terminate the core Haldane chain so that there is a Kramers doublet localized at this end point. In this case the external vortex source is a bosonic Kramers doublet. In the other example SPT (labelled $eCmC$) the vortices are ribbons with a phase factor $(-1)$ associated with each self-linking of the ribbon. Open end points of such vortex strings are fermionic Kramers singlets. Obviously stacking these two phases together produces an SPT where bulk vortex sources are fermionic Kramers doublets. In contrast in trivial boson insulators such bulk external vortex sources are bosons with trivial quantum numbers under global symmetries.

This understanding of the different SPT phases immediately determines what happens when the $U(1)$ symmetry is gauged. 
As these phases are gapped insulators (at least in the bulk) there will now be a dynamical photon. More interesting for our purposes is the fate of the magnetic monopole $m_d$. The monopole serves  as a source of $2\pi$ magnetic flux for the $e_d$ particle. Thus it should precisely be identified with the source of vortex lines. It follows that $m_d$ can therefore either be a Kramers singlet/doublet and have bose/fermi statistics. 

Reversing the duality transformation we see that these are precisely the four distinct $U(1)$ quantum liquids discussed in the previous subsection. We have thus established our promised claim that these different $U(1)$ quantum liquids may be equivalently viewed as different bosonic monopole SPT insulators.

\begin{table}[htdp]
\begin{tabular}{|c|c|}
\hline
Electric particle &  Monopole insulator  \\ \hline
$\mathcal{T}^2=1$, boson& Trivial \\ \hline
${\cal{T}}^2=-1$, boson & SPT -$eCmT$ \\ \hline
$\mathcal{T}^2=1$, fermion & SPT-$eCmC$ \\ \hline
${\cal{T}}^2=-1$, fermion & SPT- $eCTmC$=$eCmT\oplus eCmC$ \\ \hline
\end{tabular}
\caption{ Phases of $U(1)$ quantum liquids ($ Z_2^T$ symmetry and even emergent electric field), labeled by symmetry properties of the electric charge, and the corresponding type of monopole SPT, conveniently labeled by the possible surface topological order.}
\label{u1gauge}
\end{table}%

In Table \ref{u1gauge} we list all the distinct phases of the $U(1)$ gauge theory with their monopole quantum numbers, and the corresponding SPT states (labeled by the surface topological states) formed by the bosonic matter field.
Notice that SPT states descended from that of $Z_2^T$ symmetry (the $eTmT$ and the all-fermion states) didn't appear in Table \ref{u1gauge}. One can understand this by thinking of these states as combinations of trivial insulators and $Z_2^T$ SPT states formed by charge-neutral bosons, hence the $U(1)$ gauge field is decoupled from the SPT states and the vortex source ({\em i.e} monopole $m_d = e$) remains trivial.

\subsection{Odd electric field}
\label{odde}
We now consider $U(1)$ liquid states where under time reversal the electric field is odd and the magnetic field is even. In the convention of Ref. \onlinecite{hfb04} this includes the case of quantum spin ice. Again we restrict attention to $U(1)$ liquids where the magnetic monopole $m$ is bosonic and transforms trivially under $Z_2^T$. What then are the possibilities for the $e$ particle? 

Based on the insights of the previous subsection let us first see what we can learn by considering different monopole SPT phases. Now the magnetic charge $q_m \rightarrow q_m$ under time reversal so that 
\begin{equation}
{\cal T}^{-1} m^\dagger {\cal T} = m
\end{equation}
Thus $m$ (or equivalently $e_d$ after the duality transformation) transforms under $U_g(1) \rtimes Z_2^T$.  

For bosons with global symmetry $U(1) \rtimes Z_2^T$ there is one non-trivial SPT phase which is again conveniently labeled by its surface topological order $eCmC$. Other SPT phases are inherited from $Z_2^T$ and hence are not pertinent to our present concerns (see the end of Sec. \ref{gb3d}). Thus we have two possible phases - the trivial insulator and the SPT 
insulator labelled by $eCmC$. In the former case external probes where bulk  vortex lines end are bosons while in the latter they are fermions. In both cases the vortex sources are Kramers trivial.  

Let us now following the logic of the previous subsection and gauge the $U(1)$ symmetry. The resulting monopole $m_d$ will be identified with the vortex source and will therefore be a Kramers singlet which can be either boson or fermion. Thus this reasoning suggests that for the odd electric field case there are only two possibilities for the $e$ particle ($= m_d$) - it is a Kramers singlet that is either boson or fermion.

 Let's understand the above claim directly from the gauge theory point of view, independent of the argument based on SPT.  With odd electric field the electric charge at any site $q_e$ is also odd under time reversal. This implies that the $e$ particles transform under $U_g(1) \times Z_2^T$ where $U_g(1)$ is the gauge transformation generated by $q_e$.   
Notice that we have $U_{\theta}{\cal{T}}={\cal{T}}U_{\theta}$ for $U(1)\times Z_2^T$ symmetry, where $U_{\theta}$ gives the $U(1)$ rotation.Allowing for the possibility of a multi-component field $e_I$, time reversal will be implemented by 
 \begin{equation}
 {\cal T}^{-1} e_I {\cal T} = e^{-i\alpha_e} T_{IJ} e_J^\dagger
 \end{equation}
 We can always change the common phase $\alpha_e$ by defining a new time reversal operator $\tilde{\cal T} = U(\theta) {\cal T}$.  As $U(\theta)$ is a gauge transformation $\tilde{\cal T}$ and ${\cal T}$ will have the same action on all physical operators. We therefore can set $\alpha_e = 0$ (or any other value for that matter). In particular under this redefinition ${\cal{T}}^2$ goes to $(U_{\theta}{\cal{T}})^2=U_{2\theta}{\cal{T}}^2$ so that the over all phase in the action of ${\cal T}^2$ on $e$ can be changed at will, and one can always choose ${\cal T}^2=1$. The algebraic structure of $U_g(1)\times Z_2^T$ still guarantees a degenerate doublet structure, but the degeneracy here is protected by $U_g(1)\times Z_2^T$ as a whole rather than by $Z_2^T$ alone as in Kramer's theorem. In particular, one can lift the degeneracy by breaking the $U_g(1)$ symmetry but still preserving time-reversal invariance, which is in sharp contrast with the Kramer's case. It is appropriate to regard the electric charge $q_e = \pm 1$ as a non-Kramers doublet. Hence with $U_{gauge}(1)\times Z_2^T$ symmetry, any  charged particle should always be viewed as time-reversal trivial. This implies that the $e$ particle is always time reversal trivial  for a $U(1)$ gauge theory where the electric field is odd, in full agreement with what we obtained from  the SPT point of view.
 
 Before concluding this section let us briefly discuss the putative $U(1)$ spin liquid in quantum spin ice from this point of view.  We have just argued that the `spinons' (in the notation of Ref. \onlinecite{hfb04}) are not Kramers doublets. If the quantum spin ice Hamiltonian has $S_z$ conservation then the spinons will generically carry fractional $S_z$.  However the realistic Hamiltonians currently proposed\cite{balentsqspice} for quantum spin ice  do not have conservation of any component of spin. Thus the ``spinons" of the possible $U(1)$ spin liquid in quantum spin ice do not carry any quantum numbers associated with internal symmetries. Their non-Kramers doublet structure is independent of whether or not the microscopic Ising spin is itself Kramers or not. Further microscopically there are two species of electric charge $e_1, e_2$ (associated with two sub lattices of the diamond lattice formed by the centers of pyrochlore tetrahedra). Time reversal should be  implemented by letting $e_1 \rightarrow e_1^\dagger, e_2 \rightarrow \pm e_2^\dagger$ so that the physical spin operator $e^\dagger_1 e_2$ transforms as appropriate with $-$ sign for a microscopic Kramers doublet spin and the $+$ sign for a non-Kramers doublet. 
 
 Finally we note that current theoretical work treats the spinons in quantum spin ice as
 bosons. This is reasonable as the electric strings connecting them are simply made up of the physical spins and do not have the ribbon structure and associated phase factors expected if they were fermions.




\section{Discussion}
\label{disc}
In this paper we studied many aspects of the realization of symmetry in highly entangled quantum phases of matter. We relied heavily on insights obtained from recent work on {\em short range entangled} symmetry protected topological phases. Despite their short range entanglement the SPT phases provide a remarkable window into the properties of the
much more non-trivial highly entangled phases. In turn the connections to the highly entangled phases enhances our understanding of SPT phases themselves. Below we briefly reiterate some of our results and their implications. 

The very existence of SPT phases emphasizes the role of symmetry in maintaining distinctions between phases of matter even in the absence of any symmetry breaking. For highly entangled states this leads to the question of whether symmetry is realized consistently in the low energy theory of such a state.  We addressed this for the example  of $2d$ gapped topological phases described by a $Z_2$ gauge theory with time reversal symmetry (and possibly a global $U(1)$ symmetry). By combining the methods of two different recent approaches\cite{hermele,Levin} to assigning symmetry to the topological quasiparticles we showed that there are consistent symmetry realizations which nevertheless are not possible in strictly $2d$ systems.  Such states were however shown to occur at the surface of $3d$ bosonic SPT phases. Conversely we provided simple arguments that if a $Z_2$ topological order can occur at the surface of a $3d$ SPT, then it is not allowed to occur in strictly $2d$ systems. 

Thus illustrates how the study of SPT surfaces can provide a very useful ``no-go" constraint on what kinds of phases are acceptable in strictly $d$-dimensional systems. If a phase occurs at the surface of a $d+1$ dimensional SPT phase (for $d > 1$) then it cannot occur with the same realization of symmetry in strictly $d$ dimensions. A nice application of this kind of no-go constraint is to the possibility of gapless vortex fluid phases proposed to exist\cite{avl} in two space dimensions in boson/spin systems with both time reversal and global $U(1)$ symmetries. Such phases were argued to exist at the surface of $3d$ SPT phases in Ref. \onlinecite{avts12} thereby strongly suggesting that they cannot exist in strictly $2d$. We sharpened this conclusion by considering a descendant $Z_2$ topological ordered phase that is obtained by pairing and condensing vortices of this putative vortex fluid. We showed that the result was a phase that cannot exist in strict $2d$ but can of course exist at the surface of $3d$ SPT. 


The study of SPT surfaces thus gives us valuable guidance in writing down legal theories of strictly $d$-dimensional systems. It thus becomes an interesting exercise to study boundary states of SPT phases in $4+1$ dimensions as a quick route to obtaining some restriction on physically relevant effective field theories of strictly $3+1$ dimensional systems. 
Quite generally the issue of consistent symmetry realization is likely related to the existence of `quantum anomalies' in the continuum field theory. For instance the surface field theory of free fermion topological insulator (or the related sigma model with the Hopf term)  are `anomalous' and require the higher dimensional bulk for a consistent symmetry-preserving regularization. For other states such as, for example,  topological quantum field theories it will be interesting if there is a useful direct diagnostic of whether the theory is anomalous
or not.

A different aspect of our results is the development of a view point on some highly entangled states as phases in which one of the emergent excitations itself is in an SPT phase. A similar result was first established in $d = 2$ in the work of Levin and Gu\cite{levingu}. We discussed how three dimensional quantum phases with an emergent $U(1)$ gauge field may be viewed as SPT phases of the magnetic monopole excitations. In the context of $3+1$-d $U(1)$ spin liquids with time reversal symmetry, whether the electric charges  are `spinons' (i.e Kramers doublet under time reversal) and whether they are bosonic or fermionic is equivalent to the different possible SPT insulating phases of the dual magnetic monopole. This new view point may potentially be useful for future studies of these interesting gapless spin liquids and their phase transitions. As we demonstrated there is a nice consistency between possible symmetry realizations in such spin liquids and the possible corresponding SPT phases. 

Finally a very interesting outcome of our results is the explicit construction of coupled layer models for $3d$ SPT phases. For all the symmetry classes discussed in Ref. \onlinecite{avts12} we provided such a construction. The strategy is to start from a layered $3d$ system where each layer is in a $Z_2$ topological ordered state that is allowed in strict $2d$.  We then coupled the layers together to confine all topological excitations in the bulk but left behind a deconfined $Z_2$ topological state at the surface. This surface topological order was shown to match the various possible such order at SPT surfaces. In particular this scheme provides an explicit construction of a $3d$ SPT state whose surface is a time reversal symmetric gapped $Z_2$ topological ordered state with three fermionic excitations that are all mutual semions. This topological order is expected to occur at the surface of a bosonic SPT state with time reversal symmetry proposed in Ref. \onlinecite{avts12} and is not currently part of the classification of Ref. \onlinecite{chencoho2011}.

We thank M.P.A. Fisher, M. Hermele, M. Levin, M.Metlitski, A. Vishwanath, and X.-G. Wen for inspirational discussions and encouragement. 
This work was supported by Department of Energy DESC-8739- ER46872. TS was also partially supported by
the Simons Foundation by award number 229736. TS also thanks the hospitality of the Physics Department at Harvard University where part of this work was done. After this  paper was completed we learnt of  Ref. \onlinecite{ashvinbcoh} which constructs a lattice model of the `beyond cohomology' 3d SPT state with $Z_2^T$ symmetry.   

\appendix
\section{$K$-matrix descriptions of $Z_2$ topological order}
\label{2dkz2}
In this appendix we consider $2d$ states in detail. In most cases a $2\times2$ $K$-matrix is enough to describe the state because we can identify all particles 
with the same symmetry and topological properties through condensing appropriate combinations of them, and there remains only one species of e and m particle, respectively. 
For example, consider a Kramer's doublet carrying spin-1/2 $b_{\pm}$, the combination $b_+b_-$ is a singlet under time-reversal and carries no spin, so we can condense it and identify 
$b_-\sim b^{\dagger}_+$, and time-reversal could be realized through $b_+\to ib_+^{\dagger}$ so that ${\cal T}^2=-1$.

The $2\times 2$ $K$-matrix was considered thouroughly in the main text, and it was straightforward to get all the possible states within the framework. It is also clear from the 
analysis above that $2\times 2$ $K$-matrix is enough to describe every state with $Z_2^T$ and $U(1)\times Z_2^T$ (spin) symmetries. For $U(1)\rtimes Z_2^T$ (charge) symmetry, the $2\times 2$ 
$K$-matrix describes most of the states, except when there is at least one particle that carries both half-charge and Kramer's doublet, in which case there is no 
particle bilinear that preserves all symmetries, and we should really consider two species of such particles. For these states, a $4\times 4$ $K$-matrix is needed.

The general form of such $K$-matrices was given in Ref. \onlinecite{Levin}, with slight modifications due to the bosonic nature of our systems here. 
There are three possible forms of $K$ and $T$ matrices. The simplest one of them
\be
K=\lp\begin{array}{cc}
      0 & A_{2\times 2} \\
      A_{2\times 2} & 0
     \end{array}\rp, T=\lp\begin{array}{cc}
                       -1_{2\times 2} & 0 \\
                        0 & 1_{2\times 2} \end{array}\rp
\ee
does not work because the $T$ matrix does not allow a particle to carry both charge and Kramer's doublet structure. The next possibility
\bea
K&=&\lp\begin{array}{cc}
      K_{2\times 2} & W_{2\times 2} \\
      W^{T}_{2\times 2} & -K_{2\times 2}
     \end{array}\rp, \tau=\lp\begin{array}{c}
                           \tau_1 \\ \tau_2 \\ \tau_1 \\ \tau_2 \end{array}\rp    \\ \nonumber
T&=&\lp\begin{array}{cc}
                       0 & 1_{2\times 2} \\
                        1_{2\times 2} & 0 \end{array}\rp
\eea
with $W_{2\times 2}$ anti-symmetric, does not work either. To see this, simply look at the charge carried by any particle $q_l=l_IK^{-1}_{IJ}\tau_J$. The entries of $K^{-1}_{IJ}$ are either integers or half-integers. From the structure of the $T$-matrix and the assumption that time-reversal doesn't interchange $e$ and $m$ particle, we find that the only half-integer entries of $K^{-1}$ are $K^{-1}_{12}=K^{-1}_{21}, K^{-1}_{14}=K^{-1}_{41}, K^{-1}_{23}=K^{-1}_{32}, K^{-1}_{34}=K^{-1}_{43}$. Then from the structure of the $\tau$ vector it is easy to see that the charge $q_l=l_IK^{-1}_{IJ}\tau_J$ must be an integer for any integer vector $l$, so there's no quasi-particle that carries half-charge.

The only possibility left is thus
\bea
K&=&\lp\begin{array}{cccc}
      0 & A & B & B \\
      A & 0 & C & -C \\
      B & C & D & 0 \\
      B & -C & 0 & -D
     \end{array}\rp, \tau=\lp\begin{array}{c}
                           0 \\ \tau_2 \\ \tau_3 \\ \tau_3 \end{array}\rp    \\ \nonumber
T&=&\lp\begin{array}{cccc}
      -1 & 0 & 0 & 0 \\
      0 & 1 & 0 & 0 \\
      0 & 0 & 0 & 1 \\
      0 & 0 & 1 & 0
     \end{array}\rp
\eea
with ${\rm det}K=(AD-2BC)^2=4$. The inverse of the $K$ matrix is thus
\be
K^{-1}=\frac{{\rm sgn}(AD-2BC)}{2}\lp\begin{array}{cccc}
      0 & D & -C & -C \\
      D & 0 & -B & B \\
      -C & -B & A & 0 \\
      -C & B & 0 & -A
     \end{array}\rp
\ee
Therefore to have the right self and mutual statistics, we need $A=4m,D=2n$, and $B,C$ odd, which makes particle-1 ($l=(1,0,0,0)$) or 2 ($l=(0,1,0,0)$) having $\pi$-statistics with particle-3 ($l=(0,0,1,0)$) or 4 ($l=(0,0,0,1)$), and all 
the other mutual or self statistics trivial.

It is clear from the $T$ matrix that particle-2 is time-reversal trivial. Since the bound state of particle-1 and particle-2 ($l=(1,1,0,0)$) has trivial statistics with any 
particle from the structure of $K^{-1}$, it must be physical hence time-reversal trivial, which implies that particle-1 should also be time-reversal trivial. Now consider the charge 
of these two particles. It is straightforward to see that with the given charge vector $\tau$, the charge carried by particle-1 or 2 
$q_{1,2}=\tau_IK^{-1}_{IJ}l_J, (J=1,2)$ can only be an integer. Hence particle-1 and 2 carry neither fractional charge nor Kramer's doublet.

Recall that our purpose here is to describe phases with a particle that carries both charge-$1/2$ and Kramer's doublet. Hence particle-3 and particle-4 must form a 
Kramer's doublet and carries charge-$1/2$. So we want the charge vector that makes $q=\tau_IK^{-1}_{IJ}l_J$ half-integer when 
$l\in\{(0,0,1,0),(0,0,0,1)\}$. It is then straightforward to show that we need 
$\tau_2$ to be odd and $\tau_3=\tau_4$ to be any integer. 

What we have shown above is that if the $e$-particle carries both charge-1/2 and Kramer's doublet structure, the $m$-particle must be trivial under both symmetry transforms, i.e. the phase has to be $eCT$.

\section{Dual Landau-Ginzburg theory of SPT surface}
\label{duallgw}
In this Appendix we briefly describe the Landau-Ginzburg theory of the surface of 3d SPT states (symmetry $U(1) \rtimes Z_2^T$ or $U(1) \times Z_2^T$) in terms of dual vortex fields. 
Ref. \onlinecite{avts12}  showed that  the difference with a trivial surface is captured very simply in terms of the difference in the structure of the vortex (this should be understood as the point of penetration of the $3d$ vortex line with the surface). Here we elaborate on this dual theory and present an explicit derivation starting from the surface topological ordered phase. 

Let us consider  $U(1) \times Z_2^T$ and consider the phase labeled by surface topological order $eCmC$ (the symmetry $U(1) \rtimes Z_2^T$ analysis is essentially the same). According to Ref. \onlinecite{avts12} the boundary vortex is then a Kramers singlet fermion. The corresponding surface Landau-Ginzburg Lagrangian may be written schematically 
\begin{equation}
\label{fvLG}
{\cal L}_d = {\cal L}[c, a_\mu] + \frac{1}{2\pi}A_\mu \epsilon_{\mu\nu\lambda} \partial_\nu a_\lambda
\end{equation}
The first term describes a (spineless)  fermionic field $c$ coupled minimally to the dual internal gauge field $a_\mu$, and $A_\mu$ is an external probe gauge field. The field $c$ describes the fermionic vortex. The global $U(1)$ current is as usual 
\begin{equation}
j_\mu = \frac{1}{2\pi}  \epsilon_{\mu\nu\lambda} \partial_\nu a_\lambda
\end{equation}
If instead the field $c$ were a boson the Lagrangian above would be the standard dual Lagrangian for a system of strictly $2d$ bosons. Let us first describe the phase structure of this dual fermionic vortex theory. We will then provide an explicit derivation that is complementary to the general considerations of Ref. \onlinecite{avts12}. 

 If the fermionic vortex $c$ is gapped and in a trivial `band' insulator, then as usual we get a surface superfluid. Note that as $c$ is a fermion it cannot condense. This precludes the usual mechanism of vortex condensation to obtaining a trivial boson insulator as expected for an SPT surface.  The surface superfluid order can be killed if pairs of $c$ condense, 
 {\em i.e} $<cc> \neq 0$.  This leads to a surface topological order described by a $Z_2$ gauge theory. There is the unpaired fermion that survives as a gapped excitation carrying zero global $U(1)$ charge. We identify this with the neutral $\epsilon$ particle. The pair condensation quantizes flux of $a_\mu$ in units of $\pi$. This  carries global $U(1)$ charge $1/2$ and we identify this with the $e$ particle. It follows that this is the $eCmC$ phase.  
 
As described in Ref. \onlinecite{avts12}  if we break time reversal at the surface we can get a gapped phase without topological order. This is obtained by simply letting the $c$-fermionic vortices completely fill a topological band with Chern number $\pm 1$, {\em i.e} the Hall conductivity of the $c$-fermion is $\sigma^c_{xy} = \pm 1$.  To see that this indeed describes the correct $T$-broken surface state we use a Chern-Simons description of this state. First rewrite the fermion current $j^f$ in terms of a dual gauge field $\tilde{a}$: 
\begin{equation}
j^f_\mu = \frac{1}{2\pi}\epsilon_{\mu\nu\lambda} \partial_\nu \tilde{a}_\lambda
\end{equation}
When the fermion has Hall conductivity, say $+1$, the effective Chern-Simons Lagrangian in terms of $(a, \tilde a)$ becomes
\begin{equation}
{\cal L} = \frac{1}{4\pi} \tilde{a} d\tilde{a} + \frac{1}{2\pi} a d\tilde a  + \frac{1}{2\pi} A da
\end{equation}
(We have used the compact notation $da \equiv \epsilon_{\mu\nu\lambda} \partial_\nu a_\lambda$). This is a 2-component Chern-Simons theory with a $K$-matrix
\be
K=\lp\begin{array}{cc}
          1 & 1 \\
          1 & 0 \end{array}\rp
\ee
with charge vector $\tau = (0, 1)$. This state has electrical Hall conductivity $\sigma_{xy} = \tau^T K^{-1} \tau = -1$. Further as this $K$-matrix has one positive and one negative eigenvalue, it is a non-chiral state with $\kappa_{xy} = 0$. Finally there is no surface topological order as $|det K| = 1$. These are exactly the right properties of the $T$-broken surface state without topological order at this SPT surface. 

Thus the fermionic vortex Landau-Ginzburg theory correctly reproduces the surface phase structure of this SPT phase. We note that if the $c$-fermion had band structure with an even 
number of gapless Dirac fermions we get exactly the proposed Lagrangian for the AVL phase, consistent with the claim in Sec. \ref{avl} that the AVL state can only occur at the surface of an SPT state (with T-reversal) and not in strict $2d$. 

A different SPT state with $U(1) \times Z_2^T$ symmetry has a bosonic Kramers doublet vortex $z_\alpha$, $\alpha = 1,2$. The corresponding dual Landau-Ginzburg theory takes the form
\begin{equation}
{\cal L}_d = {\cal L}[z_\alpha, a_\mu] + \frac{1}{2\pi} A da
\end{equation}
Under time reversal $z_\alpha \rightarrow i\sigma^y_{\alpha\beta} z_\beta$.  Finally by stacking the two SPT phases described above we obtain a third SPT with a fermionic Kramers doublet vortex field $c_\alpha$ with Lagrangian
\begin{equation}
{\cal L}_d = {\cal L}[c_\alpha, a_\mu] + \frac{1}{2\pi} A da
\end{equation}

The surface phase structure of these other SPTs can be readily discussed in terms of these dual Landau-Ginzburg theories. In all these cases there is a bulk-edge correspondence that relates the structure of the surface vortex to the properties of the bulk monopole when the global $U(1)$ symmetry is gauged.  Including the trivial (i.e non-SPT) insulator, we have four possible SPT phases with four distinct surface vortices (end points of bulk vortex lines). These correspond precisely to the four possible bulk monopoles of the gauged SPT as discussed in Sec. \ref{gb3d}. 

We now provide an explicit derivation of Eqn. \ref{fvLG} for the corresponding SPT phase.  Let us begin with the surface topological order $eCmC$.  Under time reversal the $e, m$ particles transform as 
\begin{eqnarray}
{\cal T}^{-1} e{\cal T}  & =  & e^\dagger \\
{\cal T}^{-1} m{\cal T}  & =  & m^\dagger 
\end{eqnarray}
while the $\epsilon$ particle is left invariant.  It is convenient for our purposes to focus on the $e$ (described by a boson field $b$) and $\epsilon$ (described by a fermion $f$) particles with mutual semion interactions. We will implement this in a lattice model of the surface through two Ising gauge fields $\sigma, \mu$ with a mutual Ising Chern-Simons term\cite{SenthilFisher}. The mutual Chern-Simons term imposes a constraint relating the integer valued lattice 3-current $\vec j_b$ to the Ising gauge flux of $\sigma$ that the fermion sees: 
\begin{equation}
\label{z2ms}
\left(-1\right)^{\vec j_b} = \prod_P \sigma_{ij}
\end{equation}
Here the plaquette product in the RHS is taken over the space-time plaquette pierced by the link of the dual lattice on which the boson current flows. The lattice space-time Lagrangian may be taken to be 
\begin{eqnarray}
{\cal L} & = & {\cal L}_b + {\cal L}_f  \\
{\cal L}_b & = & \kappa\vec  j_{b\alpha}^2 + i\vec  j_{b\alpha}.\frac{ \left(\vec A_\alpha \right)}{2} \\
{\cal L}_f &  = & -\sigma_{ij} \left(t_{ij} f^\dagger_i f_j + h.c + ...\right)
\end{eqnarray}
The fermion Lagrangian will in general also include pairing terms $f_i f_j + h.c$. As before $A$ is the external probe gauge field. 
We now implement a standard duality transformation on the $b$ field by first writing $\vec j_b = \vec \nabla \times \vec \alpha$ with $\alpha$ an integer.   As this kind of duality has been explained in detail in Refs. \onlinecite{SenthilFisher,tsfisher06} we will be very brief. The mutual semion constraint Eqn. \ref{z2ms} can be solved to write 
\begin{equation}
(-1)^{\alpha} = \sigma
\end{equation}
This means that the integer $\alpha = 2\alpha' + \frac{\pi}{2}\left(1 - \sigma\right)$ with $\alpha'$ an integer. Imposing the integer constraint on $\alpha'$ softly leads to a term
\begin{equation}
-\lambda cos\left(2\pi \alpha' \right) = -\lambda \sigma_{ij} cos\left(\pi \alpha_{ij}\right)
\end{equation}
We may now define $\pi \alpha =  a$ and extract a longitudinal piece $a_{ij} \rightarrow a_{ij} + \phi_i - \phi_j$ to obtain a dual vortex Lagrangian in Euclidean space-time:
\begin{eqnarray}
{\cal L} & = & {\cal L}_\phi + {\cal L}_f \\
{\cal L}_\phi & = & -\lambda \sigma_{ij} cos \left(\vec \nabla \phi + \vec a \right) + \frac{\kappa}{\pi^2}\left( \vec \nabla \times \vec a \right)^2 \\
& & + i\frac{1}{2\pi} \vec A. \vec \nabla \times \vec a
\end{eqnarray}

Finally tracing over the Ising gauge fields $\sigma$ gives a dual vortex theory in terms of two fermionic fields $c_\pm$ defined through
\begin{equation}
c_{\pm} = f e^{i\pm \phi}
\end{equation}
However generically due to pairing terms in the $f$ Lagrangian $c_-$ can mix with $c_+^\dagger$ so that there is a unique fermion field $c = c_+ \sim c_-^\dagger$. 
This then gives us the dual fermionic vortex Landau-Ginzburg theory in Eqn. \ref{fvLG}. For $U(1) \times Z_2^T$ under time reversal we must have $c \rightarrow c$. The 
dual vortex theory for the surface of the $U(1) \rtimes Z_2^T$ SPT (with $eCmC$ surface topological order) can be derived identically, and takes the same form except that under time reversal $c \rightarrow c^\dagger$.

\end{document}